\documentstyle[pra,aps,preprint]{revtex}

\def\centerwmf#1#2#3{\vspace*{#2}\relax\centerline{\hbox to#1{\special{wmf:#3 x=#1, y=#2}\hfil}}}
\def\centereps#1#2#3{\vspace*{#2}\relax\centerline{\hbox to#1{\special{eps:#3 x=#1, y=#2}\hfil}}}

\begin{document}
\draft 
\title{Generating and probing a two-photon Fock state with a single atom in a cavity}
\author{P. Bertet$^1$, S. Osnaghi$^1$, P. Milman$^1$, A. Auffeves$^1$, P. Maioli$^1$, 
M. Brune$^1$, J.M. Raimond$^1$ and S. Haroche$^{1,2}$}

\address{$^1$Laboratoire Kastler Brossel, D\'epartement de Physique de
l'Ecole Normale Sup\'erieure,\\ 24 rue Lhomond, F-75231 Paris
Cedex 05 France\\
$^2$ Coll\`ege de France, 11 place Marcelin-Berthelot, F-75005, Paris France}
\date{\today}
\maketitle

\begin{abstract}
A two-photon Fock state is prepared in a cavity sustaining a ``source mode " and 
a ``target mode", with a single circular Rydberg atom. In a third-order Raman process, the 
atom emits a photon in the target while scattering one photon from the source into the target. 
The final two-photon state is probed by measuring by Ramsey interferometry 
the cavity light shifts induced by the target field on the same atom. Extensions to 
other multi-photon processes and to a new type of micromaser are briefly discussed.
\end{abstract}
 \vskip1pc

The generation of non-classical states of light has been the subject of an intense 
theoretical and experimental activity since the first observations of squeezed 
states of light \cite{SQUEEZED}. The study of these states provides a fundamental 
understanding of quantum fluctuations and opens the way for new schemes of 
communication or imaging beating the standard quantum noise limit. Entangled states 
produced in down-conversion processes have been widely used to test fundamental 
quantum features such as non-locality \cite{ZEILRMP} or to realize quantum 
information transmission schemes (cryptography \cite{CRYPTO} or teleportation 
\cite{TELEPORT}). Quantum superpositions of fields with different classical 
parameters have been used to explore the quantum/classical boundary and the 
decoherence phenomenon \cite{EXPCAT}.

In this context, Fock or photon number states are particularly interesting. 
They exhibit no intensity fluctuations and a complete phase indetermination. 
Their Wigner function presents negativities which reveal clearly their 
non-classical nature \cite{WIGNER,MLYNEK}. Many studies have been devoted to 
their production and characterization. Single-photon Fock states are easily 
prepared from a pair of entangled photons by the detection of one of the 
members of the pair \cite{GRANGIER}. They can also be prepared ``on demand" 
by controlling the emission of a single radiator: molecule \cite{ORRIT},
colored center \cite{DIAMOND} or quantum dot \cite{QDOT}. These single-photon 
states are useful for quantum information transmission schemes \cite{CRYPTO}. 
Single-photon sources could also be an essential resource in an all-optical 
quantum information processing device \cite{OPTICALCOMP}.

Various schemes of Fock state preparation have been proposed in the context 
of Cavity Quantum Electrodynamics (CQED) \cite{BERMAN}, in which atoms 
interact one at a time with a high $Q$ resonator. A one-photon Fock state 
has been produced in this way by a $\pi$ quantum Rabi pulse in a microwave 
cavity \cite{MEMORY} or by an adiabatic passage sequence in an optical 
cavity \cite{REMPE}. With a train of atoms, arbitrary multiphoton Fock states 
can in principle be generated in a cavity by post-selecting the events in 
which each atom has emitted one photon. Two-photon states have been generated 
in this way \cite{TWOWALTHER}. The operation of c.w. micromasers can also 
lead to Fock states preparation. The micromaser field evolves into a number 
state, provided the atomic velocity is conveniently tuned (trapping states 
\cite{TRAPPING}). A one-photon Fock state has been generated and probed in 
this way \cite{TRAPPING1}. In all these CQED processes, at least one atom 
per produced photon is required. Schemes more economical in atomic resources 
have been proposed \cite{PUMP,FRANCA} but never implemented so far. 

We report in this Letter the production of a two-photon Fock state by a single 
atom in a high-$Q$ cavity. The scheme involves the non-resonant interaction of 
the atom with two non-degenerate cavity modes. The atom, in a single non-linear process, 
emits one photon in the initially empty target mode, while transferring another 
photon from the source mode (containing initially a small field) into the target. 
This third-order Raman process, resonant for proper atom-cavity mode detunings, 
is highly efficient. 

The principle of the experiment is sketched on Fig. 1(a). More details can be 
found elsewhere \cite{RMP}. Rubidium atoms effusing from oven $O$ are 
velocity-selected and promoted, one at a time, in the circular state $e$ 
(principal quantum number 51) in box $B$. The atom in the circular state crosses 
the superconducting cavity $C$, which sustains two non-degenerate orthogonally 
polarized modes $M_a$ and $M_b$. They have the same gaussian geometry with a 
$w=6$ mm waist. Their damping times are 1.2 ms for $M_a$ and 0.9 ms for $M_b$. 
The $M_a-M_b$ frequency difference is $\delta/2\pi=128$ kHz ($M_a$ has the 
highest frequency). At thermal equilibrium, $M_a$ and $M_b$ contain a thermal 
field with one photon on average. A field ``erasure" procedure can be applied 
to prepare both modes in the vacuum state $|0\rangle$ \cite{RMP}. An external 
source $S$ can be used to inject in $M_b$ a small coherent field with a 
tunable photon number. The modes are nearly resonant with the transition 
between circular states $e$ and $g$ (principal quantum number 50) at 51.1 GHz. 
The detuning $\Delta$ between the atomic frequency and $M_a$ can be tuned by 
changing the cavity length (slow tuning) or by Stark-shifting the atomic 
line (fast tuning). The final state of the atom is analyzed by the 
state-selective field-ionization detector $D$.

An atom entering an empty cavity in state $e$, resonant with $M_a$ ($\Delta=0$) 
or $M_b$ ($\Delta=-\delta$), would undergo a quantum Rabi oscillation at the 
frequency $\Omega/2\pi=49$ kHz \cite{QRABI}. We consider here a situation 
where the atomic transition is tuned above the frequency of mode $M_a$ 
($\Delta>0$). Direct photon emission in the cavity mode is thus forbidden by 
energy conservation, provided $\Delta$ is much larger than $\Omega$. 

A higher order emission process in the initially empty target mode $M_a$ 
is however possible when the source mode $M_b$ initially contains a small 
field with a photon number distribution $p(n)$. Let us first consider 
the contribution of the $n$-photon Fock state in $M_b$. The initial 
system's state, taken as the energy reference, is $|e,0,n\rangle$. The 
symbols in the ket refer to the atomic state and to the $M_a$ and $M_b$ 
photon numbers successively. The initial state is coupled by photon 
emission in $M_a$ to $|g,1,n\rangle$ (energy $-\hbar\Delta$) [see level 
scheme on Fig. 1(b)]. This level is coupled by a one-photon absorption 
in $M_b$ to $|e,1,n-1\rangle$ (energy $\hbar\delta$), coupled in turn 
to the final state $|g,2, n-1\rangle$ (energy 
$\hbar(\delta-\Delta)$). When $\Delta=\delta$, this third-order Raman 
process is resonant and injects two photons in $M_a$ using a single atom. 
Note that a symmetrical Raman process emits two photons in $M_b$ when 
$M_a$ contains a small field and $\Delta=-2\delta$. These Raman processes
are related to the cascade in the dressed atom levels used in \cite{MOSSBERG}
to realize a two-photon optical laser oscillator with macroscopic atom
samples. We are, however, interested here in a single step process
involving only one atom.

The lowest order Raman coupling matrix element is 
$\Omega^3\sqrt{2n}/[8\Delta(\Delta+\delta)]$. The $\sqrt{n}$ and $\sqrt{2}$ factors 
arise from the stimulation of the $|g,1,n\rangle\rightarrow |e,1,n-1\rangle$ 
and of the $|e,1,n-1\rangle\rightarrow |g,2, n-1\rangle$ couplings by the 
fields in $M_b$ and $M_a$ respectively. For a coherent or thermal source field, 
the two-photon emission probability is obtained by summing independently the various 
Fock state contributions in $M_b$, weighted by their probability $p(n)$. 
The efficiency of the process thus increases with the average photon number 
in the source mode. 

This simple discussion neglects the light shifts produced by the non-resonant 
modes on the atomic levels. These second order effects are much larger than
the Raman coupling.
An effective hamiltonian model and a 
numerical simulation of the Raman emission process \cite{TOBE} show that 
these shifts reduce the $\Delta$ value at resonance, 
which becomes smaller than $\delta$ and decreases with the photon number 
in $M_b$ (for instance, for a six photon coherent field in $M_b$, the 
resonance condition is $\Delta/2\pi=65$ kHz). For large source fields, 
the Raman transition merges with the 
ordinary emission line at $\Delta=0$.

Figure 2 presents the probability $P_g$ for detecting the atom in state 
$g$, reconstructed by averaging 1000 realizations of the experiment, as 
a function of $\Delta$. The atomic velocity is $v=200$ m/s. The solid 
squares on figure 2(a) (with statistical error bars) correspond to both 
modes initially at thermal equilibrium. The main features are the two 
direct one-photon emission resonances at $\Delta=0$ and $\Delta=-\delta$. 
Two Raman resonances are also observed around $\Delta/2\pi=80$ kHz and
$\Delta/2\pi=-210$ kHz. For the first, $M_a$ is the target mode 
(see discussion above). For the second, $M_a$ is the source and 
$M_b$ the target mode. These 
features are reproduced by the numerical simulation of the evolution 
(solid line). The Raman process in this situation is produced by the 
thermal field in the source mode and enhanced by the initial thermal 
field in the target mode. These Raman resonances are not observed with 
fast atoms (above 300 m/s). The $\Delta=0$ and $\Delta=-\delta$ lines 
are wider in this case and the Raman transfer rate is much smaller. The 
constant 8\% background in figure 2 is mainly due to spontaneous 
emission outside the cavity, enhanced by the background thermal field. 

The other data sets in figure 2(a) correspond to an initially empty 
$M_a$, $\Delta$ being scanned only around the upper Raman resonance. Open squares, 
obtained when $M_b$ is empty, exhibit no resonant transfer. This 
demonstrates the essential role of the source field in the Raman process. 
The two other data sets correspond to a coherent field in $M_b$ with 
an average photon number of 5.6 (open circles) and 11.2 (open squares). 
This field has been independently calibrated by a procedure described 
in \cite{COMPLEM}. These last two data sets are enlarged on figure 2(b), 
together with the results of the numerical simulations (dotted and solid 
lines). The simulations agree reasonably well with the experiment 
(note that the effect of the light shifts on the resonance frequency is here clearly 
apparent). For the largest source field, the transfer rate peaks 
at 30\%. For higher photon numbers, the Raman resonance merges into 
the one-photon line. 

We have also directly probed the final field in $M_b$. The same atom 
undergoes the Raman emission process and measures the field through 
the dispersive light shift experienced by the atomic level $g$ 
\cite{FROMLAMB}. The atom is initially prepared in $e$, with a velocity 
$v=170$ m/s. Mode $M_a$ is empty and $M_b$ contains a coherent field 
with 6 photons on the average. The cavity is initially tuned to the 
Raman resonance condition ($\Delta/2\pi=65$ kHz). The two-photon 
emission builds up. The frequency detuning $\Delta/2\pi$ is suddenly 
increased from 65 kHz to 135 kHz, at a time $t=5\ \mu$s after the atom 
has crossed the cavity axis. The Raman process is frozen from then on. 
The atom then undergoes a classical $\pi/2$ pulse which prepares a 
superposition of $g$ and $i$ (circular state with principal quantum 
number 49). This pulse, resonant on the $g\rightarrow i$ transition 
at 54.3 GHz, is fed from a classical source $S'$ into a low-$Q$ transverse 
mode of the cavity structure. The phase of the atomic 
superposition is sensitive to the light shifts produced on $g$ by $M_a$ 
and $M_b$ (the $g\rightarrow i$ transition being far off-resonant with 
$M_a$ and $M_b$, the light shifts of level $i$ is negligible). These 
light shifts are proportional to the photon number in each mode and 
inversely proportional to the final values of $\Delta$ and $\Delta+\delta$. 
The final atomic superposition is probed by applying on the atom, after 
it exits $C$, a second $\pi/2$ pulse on the $g\rightarrow i$ transition, 
resulting in a Ramsey fringe pattern \cite{RMP}. The probability for 
detecting the atom in $g$ is sinusoidal versus the frequency of $S'$. 
The phase of this modulation reflects the accumulated light shift 
\cite{FROMLAMB}. Note that the final detection of the atom in $g$ or 
$i$ selects the events in which the Raman emission has taken place. 
If the atom is in $e$ at time $t$ (no Raman emission has occurred), 
it is unaffected by the Ramsey pulses. These events are discarded.

The corresponding fringes are plotted on figure 3 (solid squares), 
compared to two reference signals. The open square signal is obtained 
when $M_a$ is empty ($\Delta$ is detuned from the Raman resonance at 
all times). The phase of these fringes is defined as being zero. The 
open diamond signal corresponds to a one-photon Fock state in $M_a$. 
The Raman emission is replaced in this case by a resonant $\pi$ Rabi 
pulse in $M_a$. Note that, for the three data sets, $M_b$ initially
contains the six photon coherent field.
The phase of the fringes obtained after the Raman 
process and after the single photon emission are $\phi_2=0.52\pm 0.04$ 
rd and $\phi_1=0.32\pm 0.03$ rd respectively. The zero phase corresponds 
to the light shifts produced by the vacuum in $M_a$ and the 6 photon 
coherent field in $M_b$. The phase $\phi_1$ reflects the apparition 
of one photon in $M_a$ while $M_b$ remains unchanged. The phase 
$\phi_2$ results from the simultaneous change of fields in $M_a$ and 
$M_b$. The first field increases by two photons, thus producing a 
shift $2\phi_1$, and the second decreases by about one, thus reducing 
the phase by about $\phi_1/2$, since $\Delta+\delta\simeq 2\Delta$. 
With the precise detuning values, we predict $\phi_2/\phi_1=1.52$, 
in good agreement with the measured ratio $1.62\pm 0.05$. 

For a more quantitative analysis, we perform a numerical simulation 
of the experiment taking into account all known imperfections of the 
set-up: cavity relaxation, spontaneous emission events before or after 
the interaction with $C$, finite contrast of the Ramsey interferometer, 
two-atom events (the atom number has a Poisson distribution with a low 
average value \cite{RMP}). The lines in figure 3 result from this simulation. 
They agree well with the experimental data. This confirms our interpretation
of the emission process.

The results of this simulation can be used to determine the photon number
distribution after the Raman emission process, at time $t$. The raw probability
for having two photons is 37\%. The broadening of the distribution is due
to two main experimental imperfections. The $e\rightarrow g$ spontaneous emission
before the atom enters the cavity generates a 20\% background of events 
where the atom enters $C$ in $g$. This imperfection could easily be removed by 
emptying level $g$ immediately before the atom enters $C$ with a convenient 
microwave pulse, raising the two-photon probability up to 46\%. The second 
imperfection is the growth of a one-photon residual thermal field in the 
cavity during the experiment. This background could be suppressed easily
(the corresponding experimental improvements are in progress). The two-photon
probability would raise then to 63\%.

We have shown in this Letter that a single atom can prepare efficiently 
a two-photon Fock state in a cavity mode by a third-order Raman process. 
This opens the way to interesting extensions. A coherent superposition of zero- and
two-photon Fock states can be readily obtained by preparing the atom intially
in a superposition of $e$ and $g$. We have also observed, for 
$\Delta=\delta$, the reverse Raman process coupling $|g,0,n\rangle$ to 
$|e,1,n-2\rangle$. The atom leaves a photon in $M_a$ and is simultaneously 
excited. This process removes two photon at the same time from the source 
mode $M_b$. A stream of atoms performing this Raman process could thus be 
used to implement an effective two-photon relaxation in mode $M_b$, 
opening interesting perspectives for relaxation tailoring and quantum states engineering.  

With a continuous source of atoms, these Raman processes lead to a new 
type of micromaser \cite{BERMAN}. We are now exploring theoretically 
its properties. It exhibits novel features, compared to the one- and 
two-photon micromasers \cite{BERMAN,TOBE} or to the two-photon
lasers \cite{MOSSBERG}. Fock states with higher photon 
numbers can also be reached. When $\Delta=2\delta$, a fifth order process 
resonantly couples $|e,0,n\rangle$ to $|g,3,n-2\rangle$, injecting three
photons in a single step into the target mode. With the atomic 
velocities available now, the corresponding transfer rate is too low 
(below $10^{-3}$) to be observed. The use of laser cooling techniques 
to prepare very slow circular atoms could allow the observation of this effect.

{\sl Acknowledgements}~: Laboratoire Kastler Brossel is a laboratory of
Universit\'e Pierre 
et Marie Curie et ENS, associated to CNRS (UMR 8552). We acknowledge support 
of the European Community, of the Japan Science and Technology corporation 
(International Cooperative Research Project~: Quantum Entanglement). 
P. Milman acknowledges support by the Brazilian agency CNPq.

\begin{figure}
\centereps{7.5cm}{9.2cm}{scheme.eps}
\caption{(a) Scheme of the experiment. (b) Energy diagram for the 
two-photon Raman emission.} \label{FIG1}
\end{figure}

\begin{figure}
\centereps{7.5cm}{9cm}{transfer.eps}
\caption{Probability $P_g$ for detecting the atom in $|g\rangle$ as a 
function of $\Delta$. (a) Solid squares: both cavity modes contain 
initially a thermal field. Error bars reflect the binomial detection 
statistics variance. Spontaneous emission resonances are observed at $\Delta=0$ 
and $\Delta=-\delta$. The two Raman resonances are around 
$\Delta/2\pi=80$ kHz and $\Delta/2\pi=-210$ kHz. The solid line results from 
a numerical simulation. Open squares: both modes initially empty. 
Open circles and diamonds: $M_a$ initially empty and $M_b$ 
ontaining 5.6 and 11.2 photon coherent fields respectively. 
(b) Enlargement of the Raman resonances for 5.6 (open squares) and 
11.2 (open diamonds) photons in $M_b$. Lines result from a numerical simulation.
} \label{FIG2}
\end{figure}

\begin{figure}
\centereps{7.5cm}{5.4cm}{fringes.eps}
\caption{Ramsey interference fringes on the $g\rightarrow i$ transition 
for three different states of the field in $M_a$. Open squares: 
reference fringes for $M_a$ in the vacuum state. Open diamonds: 
$M_a$ contains a single photon Fock state produced by a $\pi$ 
resonant spontaneous emission pulse. Solid squares: $M_a$ state 
results from a Raman transition. The lines are provided by a 
numerical simulation of the experiment (see text).} 
\label{FIG3}
\end{figure}

\end{document}